\documentstyle[prc,aps,preprint,psfig]{revtex}
\title{Large Bipolaron in a Polaron-Gas Background}
\author{M.~A.~Smondyrev\cite{d}, A.~A.~Shanenko\cite{d},
J.~T.~Devreese\cite{x1}}
\address{Universiteit Antwerpen (UIA), Departement
Natuurkunde, \\
Universiteitsplein 1, B-2610 Antwerpen, Belgium}
\begin{document}
\maketitle
\pagenumbering{arabic}
\begin{abstract}
A criterion, proposed by the present authors, is used to derive
numerical results for the stability of a large bipolaron
embedded in a polaron gas. The main conclusion is that an isolated
metastable bipolaron can be stabilized by the polaron gas surroundings
because of the Fermi statistics of polarons.
On the other hand, it is found that the exchange interaction
tends to destabilize the bipolaron.
The study is performed both for bulk (3D) materials
and for thin (2D) films within the Hartree-Fock approximation.
The bipolaron is described by an extension of the Feynman
polaron model.
\end{abstract}
\narrowtext

Two electrons (or holes) in a polar crystal
interact with the phonon field
which may lead to their binding into a composite
quasi-particle, a {\em bipolaron}.
The binding into a bipolaron results if the
attraction between the electrons, due to the virtual phonon exchange,
is sufficiently strong to overcome the electron-electron repulsion.

In this letter we deal with the so called large (bi)polarons
(e.g. Refs. \cite{devr,adam,iadon}; for more references see the review
articles~\cite{fomin,jtd}).
The singlet large bipolaron is characterized by
two dimensionless coupling constants. The first one, the Fr\"ohlich
coupling constant
\begin{eqnarray}
\alpha = {1 \over \hbar \omega_{\mbox{\tiny LO}}} {e^2 \over
\sqrt{2}}
\left({1\over \varepsilon_{\infty}} - {1\over \varepsilon_0}\right)
\sqrt{{m_b\omega_{\mbox{\tiny LO}}\over \hbar}}
\label{eq01}\end{eqnarray}
is a measure of the strength of the electron-phonon interaction
while the second one
\begin{eqnarray}
U = {1 \over \hbar \omega_{\mbox{\tiny LO}}} {e^2 \over
                                      \varepsilon_{\infty}}
                   \sqrt{{m_b\omega_{\mbox{\tiny LO}}\over \hbar}}
\label{eq02}\end{eqnarray}
is the Coulomb potential coupling constant which governs the
strength of the direct electron-electron repulsion. In
(\ref{eq01}) and (\ref{eq02}) $ \varepsilon_0 $ and
$ \varepsilon_{\infty} $ are the static and the high-frequency
dielectric constants, $m_b$ is the electron band mass and
$\omega_{\mbox{\tiny LO}}$ is the frequency of the longitudinal
optical (LO) phonons. Introducing the ratio of the
dielectric constants $\eta = \varepsilon_{\infty}/\varepsilon_0,$
one obtains the following relation between the Coulomb and
the electron-phonon coupling constants
$U =  {\sqrt{2}\,\alpha /(1-\eta)}$. As $\eta \geq 0$, only values
$U \geq \sqrt{2}\,\alpha$ have a physical meaning. In what follows
use is also made of the parameter
\begin{eqnarray}
u={U \over \sqrt{2}\alpha} = {1 \over 1-\eta}\ .
\label{eq02a}\end{eqnarray}
The physical relevant region is then defined by the
inequality $u \geq 1$.

To find the stability region of the bipolaron in the
($\alpha,u$)-plane one usually exploits the following condition
\begin{eqnarray}
E_{bip} \leq 2 E_{pol},
\label{eq03}\end{eqnarray}
where $E_{pol}$ and $E_{bip}$ denote the ground state energies
of the polaron and the bipolaron at rest, respectively.
This inequality implies that the decay of the bipolaron into two
polarons is not energetically advantageous.
With (\ref{eq03}) the bipolaron turns out to be stable
if $\alpha$ is larger than some critical coupling constant $\alpha_c$.
In three dimensions (3D)
the following estimates
have been reported for the critical value of the electron-phonon
coupling constant: $\alpha_c=7.3$ in Ref.~\onlinecite{adam},
$\alpha_c=6.8$ in Ref.~\onlinecite{devr} and $\alpha_c=6$ in
Ref.~\onlinecite{iadon}. Even if $\alpha$ exceeds the critical
value $\alpha_c$, the possibility of large bipolaron formation depends on
the strength of the direct Coulomb repulsion which tends to prevent
electrons to congregate into a
cluster. Namely, $E_{bip}$ obeys (\ref{eq03}) at $\eta \leq
\eta_c(\alpha)$ (or, equivalently, $u \leq u_c(\alpha)$).
For example, for large coupling, $\alpha=9$, it has been found that
$\eta_c \approx 0.056$ ($u_c \approx 1.059$) in Ref.~\onlinecite{adam}
and $\eta_c \approx 0.037$ ($u_c \approx 1.038$) in Ref.~\onlinecite{devr}.
(The quoted results for $\alpha_c$ and $\eta_c,\ u_c$ leave little hope
that the 3D singlet large bipolaron can survive even in
special kinds of materials (such as strongly ionic crystals)).

In reality,
the inequality (\ref{eq03}) is the stability criterion for an
{\it isolated bipolaron}; it is not applicable for a bipolaron
interacting with a system of charge carriers. In Ref.~\onlinecite{shan}
the present authors have proposed a stability criterion for
a 3D bipolaron embedded in a polaron gas.
The basic idea is the following. Polarons are fermions and
obey the Pauli principle. Therefore the two polarons
in a final state after the bipolaron decay must have their momenta
outside the Fermi-surface. Consequently, the total kinetic energy
of two such final state polarons cannot be less than
$2(p_F^2/2m_{pol})$ where $m_{pol}$ is the polaron effective mass.
We must add this term to the r.h.s of the criterion (\ref{eq03}). This
evidently makes the bipolaron decay less probable.
That is, the bipolaron is possibly stabilized because of the Fermi
statistics of the final state polarons.
Furthermore, the polaron-polaron and the bipolaron-polaron interactions
should be taken into account.

In the present letter we restrict ourselves to the Hartree-Fock
approximation and perform the calculations for an arbitrary number $D$
of dimensions to obtain the following formulae for the 2D bipolaron.
The $D$-dimensional Fermi-momentum is given by the expression
\begin{eqnarray}
p_F = 2\sqrt{\pi}\hbar \biggl[{n\over 2}\Gamma(1+D/2)\biggr]^{1/D},
\label{eq04}\end{eqnarray}
where $n=N/V$ is the polaron concentration in a $D$-dimensional
box. In the Hartree-Fock approximation the mean kinetic energy
per particle (without polaron effects) reads as follows
\begin{eqnarray}
W(n) = \langle {p^2\over 2m_{pol}}\rangle =
{2\pi\hbar^2\over m_{pol}}\,{D\over D+2}\,\biggl[{n\over
2}\Gamma(1+D/2)\biggr]^{2/D}=
{D\over D+2}\,{p_F^2\over 2m_{pol}}.
\label{eq05}\end{eqnarray}
A term to be added to the r.h.s of the criterion (\ref{eq03})
is the difference between the kinetic energies of $N+2$ polarons
(after the bipolaron decay) and that of $N$ polarons
(the bipolaron is assumed to be at rest):
\begin{eqnarray}
\Delta_{kin} &=& \left.(N+2)\langle {p^2\over 2m_{pol}}\rangle\right|_{N+2}
- \left. N\langle {p^2\over 2m_{pol}}\rangle\right|_{N} \nonumber \\[3mm]
&=& V \left[(n+{2\over V})W\left(n+{2\over V}\right)
-nW\left(n\right)\right]= 2 {\partial\over \partial n}\left[nW(n)\right]
\nonumber \\[3mm]
&=& {4 \pi\hbar^2\over m_{pol}} \biggl[{n\over 2}\Gamma(1+D/2)\biggr]^{2/D}
= 2\,{p_F^2\over 2m_{pol}}.
\label{eq06}\end{eqnarray}
Thus, $\Delta_{kin}$ is twice the polaron kinetic energy
on a Fermi-surface.

In a similar way we treat the potential energy of the polaron gas.
At large distances $r \gg r_{pol}$ (where $r_{pol}$ is the polaron radius)
the polaron-polaron and the bipolaron-polaron
interactions can be approximated by Coulomb potentials screened
by the static dielectric constant:
$\Phi_{pp} \sim {e^2/\varepsilon_0 r}$ and $\Phi_{bp} \sim
{2e^2/\varepsilon_0 r}$.
In the Hartree-Fock approximation the exchange energy per particle
is then approximated as follows
\begin{eqnarray}
\Pi(n) &=& -{\pi e^2\over \varepsilon_0 n}\,{\Gamma\left({D-1\over 2}\right)
\over (2\pi^{3/2}\hbar)^{D+1}}\,\int\limits_{|\vec k|,|\vec q|\leq p_F}
{d^D {\vec k}\, d^D {\vec q}\over |\vec k-\vec q|^{D-1}}.
\label{eq07}\end{eqnarray}
The integral in (\ref{eq07}) can be evaluated
\begin{eqnarray}
\int\limits_{|\vec k|,|\vec q|\leq p_F}
{d^D {\vec k}\, d^D {\vec q}\over |\vec k-\vec q|^{D-1}} =
p_F^{D+1}\, {4\pi^{(D-1/2)}\over \Gamma(D/2)\Gamma[(D+3)/2]},
\label{eq08}\end{eqnarray}
and we obtain from Eq. (\ref{eq07})
\begin{eqnarray}
\Pi(n) = -{e^2\over \varepsilon_0}\,{4D\over \sqrt{\pi} (D^2-1)}
\biggl[{n\over 2}\Gamma(1+D/2)\biggr]^{1/D}.
\label{eq09}\end{eqnarray}
The exchange energy to be added to the r.h.s of the criterion
(\ref{eq03}) takes therefore the form
\begin{eqnarray}
\Delta_{exc} = 2 {\partial\over \partial n}\left[n\Pi(n)\right]=
-{e^2\over \varepsilon_0}\,{8\over \sqrt{\pi}(D-1)}\,
\biggl[{n\over 2}\Gamma(1+D/2)\biggr]^{1/D}=
- {e^2\over \varepsilon_0}\,{4\over \pi (D-1)}\,{p_F\over \hbar}.
\label{eq10}\end{eqnarray}

Finally, in the Hartree-Fock approximation,
we arrive at the following {\em in-medium} criterion
for the bipolaron stability
\begin{eqnarray}
E_{bip} - 2E_{pol} &\leq & {p_F^2\over m_{pol}}
- {e^2\over \varepsilon_0}\,{4\over \pi (D-1)}\,{p_F\over \hbar}.
\label{eq11}\end{eqnarray}
The r.h.s. of (\ref{eq11})
is equal to zero for $p_F=0$ and for $p_F=p_d$,
where $p_d$ is given by the relation
\begin{eqnarray}
p_d = {m_{pol} e^2\over \hbar \varepsilon_0}\,{4\over \pi (D-1)}.
\label{eq12}\end{eqnarray}
In the interval $0 < p_F < p_d$, that is at small polaron concentrations,
the r.h.s. of the criterion (\ref{eq11}) is negative which allows the
bipolaron to decay even if it is stable when isolated.
In this case the exchange energy dominates and a ``stable" bipolaron
(in the sense of the criterion (\ref{eq03}))
is destabilized by the polaron gas.
In the interval
$p_F > p_d$ the effect of the Fermi statistics dominates
and the r.h.s. of (\ref{eq11}) is positive thus
additionally stabilizing the bipolaron in comparison with what
follows from (\ref{eq03}).

At $D=3$ we arrive at the stability criterion\cite{shan}
\begin{eqnarray}
E_{bip} - 2E_{pol} \leq {\hbar^2\over m_{pol}}\,
                                               (3\pi^2 n)^{2/3} -
2\,{e^2\over \varepsilon_0}\,\left({3n\over \pi}
                                         \right)^{1/3}.
\label{eq13}\end{eqnarray}
In dimensionless units the criterion (\ref{eq13}) takes the form
\begin{eqnarray}
{E_{bip} - 2 E_{pol} \over  \hbar\omega_{\mbox{\tiny LO}}} \leq
{m_b \over m_{pol}}\,\left(3\pi^2 {n\over n_0}\right)^{2/3} -
2\sqrt{2}\,(u-1)\,\alpha \,\left({3\over \pi}\,{n\over n_0}
                                                     \right)^{1/3}.
\label{eq14}\end{eqnarray}
Here $n_0$ denotes the natural unit for the polaron concentration
\begin{eqnarray}
n_0 = \left({m_b \omega_{\mbox{\tiny LO}} \over \hbar}\right)^{3/2}.
\label{eq15}\end{eqnarray}
Its typical value can be estimated with the data of
Ref.~\cite{karth}. For instance, for RbCl $m_b=0.432m_e$
($m_e$ is the electron mass) and $\omega_{\mbox{\tiny LO}}=3.4
\cdot 10^{13}\ {\rm s}^{-1}$, so $n_0 \sim 4.5 \cdot 10^{19} \
{\rm cm}^{-3}$.

For quasi-particles confined to a two-dimensional ($D=2$) layer
(\ref{eq11}) takes the form
\begin{eqnarray}
E_{bip} - 2E_{pol}\leq {\hbar^2\over m_{pol}}\,2\pi \sigma -
4\,{e^2\over \varepsilon_0}\,\sqrt{2\sigma\over \pi},
\label{eq16}\end{eqnarray}
where we used the notation $\sigma$ for the surface concentration
of polarons. Eq. (\ref{eq16}) can be written as follows
\begin{eqnarray}
&& {E_{bip} - 2 E_{pol} \over  \hbar\omega_{\mbox{\tiny LO}}}  \leq
{m_b \over m_{pol}}\,2\pi {\sigma\over \sigma_0} -
8\,(u-1)\,\alpha \,\sqrt{{\sigma\over \pi\sigma_0}}, \nonumber \\[2mm]
&& \sigma_0 = {m_b \omega_{\mbox{\tiny LO}} \over \hbar} = n_0^{2/3}.
\label{eq17}\end{eqnarray}

We used the criteria (\ref{eq13}), (\ref{eq14}) in Ref.~\onlinecite{shan}
to study the possibility of stabilizing the bipolaron, restricting
ourselves to the limiting case $u=1$ (that is, omitting the exchange
energy which diminishes the bipolaron stability region).
The goal of the present letter is to study both stabilization and
destabilization
effects in more detail using the criteria (\ref{eq14}), (\ref{eq17}).
To calculate the bipolaron and the polaron ground state energies
use is made of the bipolaron model of Ref. \onlinecite{devr}
which is a generalization of the Feynman model for the single polaron.
As for any Gaussian approximation, the results
for the polaron ground state energy and the effective mass in
different dimensions are linked to each other by
a scaling relation introduced in Ref. \onlinecite{wu},
\begin{eqnarray}
E_{pol}^{(2D)}(\alpha) = {2\over 3}\,E_{pol}^{(3D)}\left({3\pi\over
4}\alpha\right),
\quad
m_{pol}^{(2D)}(\alpha) = m_{pol}^{(3D)}\left({3\pi\over 4}\alpha\right).
\label{eq18}\end{eqnarray}
A similar scaling relation is valid for the bipolaron energy\cite{devr}
(the Coulomb coupling constant is not scaled).
At the ``destabilizing" concentration $n_d$ corresponding to the value
$p_d$ of Eq. (\ref{eq12}), the r.h.s. of the stability criterion
equals zero. The bipolaron energy then equals twice the polaron
energy. This occurs  at the boundary of the bipolaron stability
region found with the old criterion (\ref{eq03}).
In 3D we obtain $n_d$ from (\ref{eq14})
\begin{eqnarray}
{n_d\over n_0} = {16\sqrt{2} \over 3\pi^5}\left[{m_{pol}\over m_b}\,\alpha\,
\left(u_c(\alpha)-1\right)\right]^3
\label{eq19}\end{eqnarray}
and in 2D from (\ref{eq17})
\begin{eqnarray}
{\sigma_d\over \sigma_0} = {16\over \pi^3}\left[{m_{pol}\over m_b}\,\alpha\,
\left(u_c(\alpha)-1\right)\right]^2.
\label{eq20}\end{eqnarray}
To find a simple estimate for $n_d$ we may use the interpolation
formula derived in Ref. \onlinecite{smond} for the
bipolaron model used here in 3D:
\begin{eqnarray}
u_c(\alpha) = 1.08525\,{\alpha^2-10.925 \over \alpha^2 -7.969}
\label{eq21}\end{eqnarray}
(note that $u$ of the present letter differs by a factor $\sqrt{2}$ from
$u$ of the paper\cite{smond}).
Inserting Eq. (\ref{eq21}) into Eq. (\ref{eq19}) and using the value
of the polaron mass calculated in the scope of the Feynman model,
we may estimate $n_d$. For instance, at $\alpha =7$ we obtain
$m_{pol}/m_b=14.4$ and $n_d=0.01n_0$. The strong dependence of $n_d$
on $\alpha$ is mostly due to the strong dependence of $m_{pol}$ on it.
That is, the smaller is $\alpha$ the less is the bipolaron destabilization
effect.

The interpolation formula for $u_c(\alpha)$ in 2D can be found
with the same scaling law (\ref{eq18}): $u_c^{(2D)}(\alpha) =
u_c^{(3D)}(3\alpha/4\pi)$. It follows then for the 2D case
\begin{eqnarray}
u_c(\alpha) = 1.08525\,{\alpha^2-1.968 \over \alpha^2 -1.435}.
\label{eq22}\end{eqnarray}
Because of the scaling relations the value $\alpha = 4\times 9/3\pi = 3.82$
in 2D produces the same polaronic effects as $\alpha=9$ in 3D.
For this $\alpha$ the polaron mass takes the value $m_{pol}/m_b =62.7$
and Eq. (\ref{eq20}) gives the surface concentration
$\sigma_d \approx 50\sigma_0$.
The numerical solution leads at the value $\sigma_d \approx 43\sigma_0$.
For $\alpha=2.97$ in 2D, which corresponds to $\alpha=7$ in 3D,
we obtain $\sigma_d \approx 0.04 \sigma_0$.

This means that the bipolaron which is stable when isolated can decay
into two polarons for concentrations $n< n_d$ ($\sigma < \sigma_d)$.
In the opposite case $n > n_d$
($\sigma > \sigma_d$) our  in-medium stability criterion
allows the bipolaron to exist even if its
energy exceeds twice the polaron energy.
Such a bipolaron being isolated would decay, therefore we call
it a metastable bipolaron. For a given value of $\alpha$
a metastable bipolaron is seen in
the range of the Coulomb coupling constant
$u_c(\alpha) \leq u \leq u_{max}(\alpha)$.
Only a state of two free polarons was seen at $u > u_{max}(\alpha)$.

\begin{figure}
\phantom{a}\vspace*{2cm}
\hspace*{1cm}
\psfig{figure=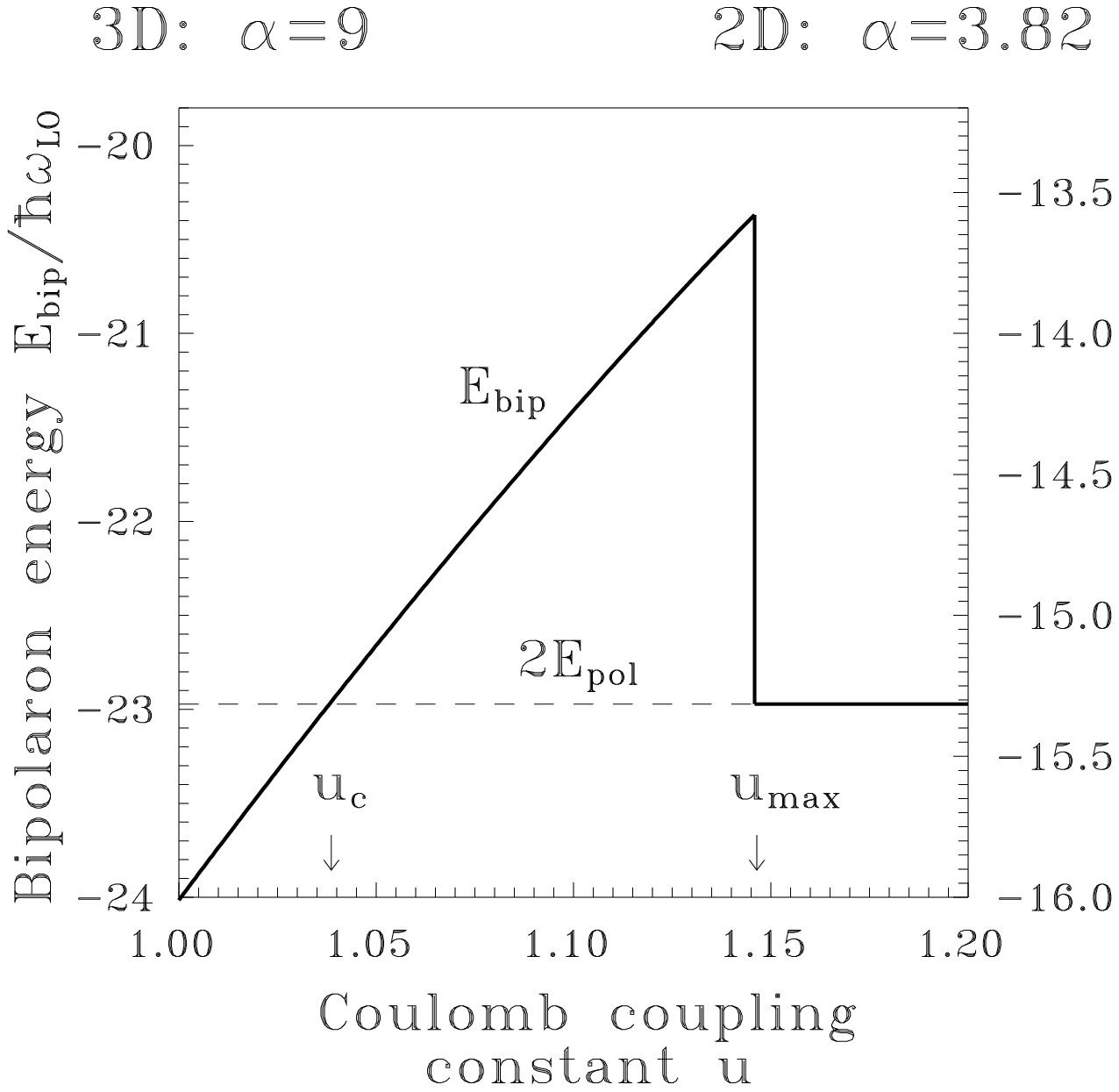,width=13cm,height=14cm}
\vspace*{1cm}
\caption{The metastable bipolaron energy vs.
the Coulomb coupling constant at $\alpha=9$ (in 3D)
and $\alpha=3.82$ (in 2D).}
\label{fig1}\end{figure}

In Fig.~\ref{fig1} the bipolaron ground state energy is
shown as a function of the Coulomb coupling constant $u$ for $\alpha=9$
in 3D. Subsequently, the same curve is obtained in 2D for
$\alpha=4\times 9/(3\pi) = 3.82$ (cf. the right $y$-axis).
In this case the isolated bipolaron is stable for $1\leq u \leq u_c=1.038$.
In the region $u_c \leq u \leq u_{max} = 1.146$
the bipolaron energy exceeds twice the
polaron energy but this metastable state can be stabilized
by the polaron gas. At $u \geq u_{max}$
our numerical program has found only two separate polarons minimum.

\begin{figure}
\hspace*{1cm}
\psfig{figure=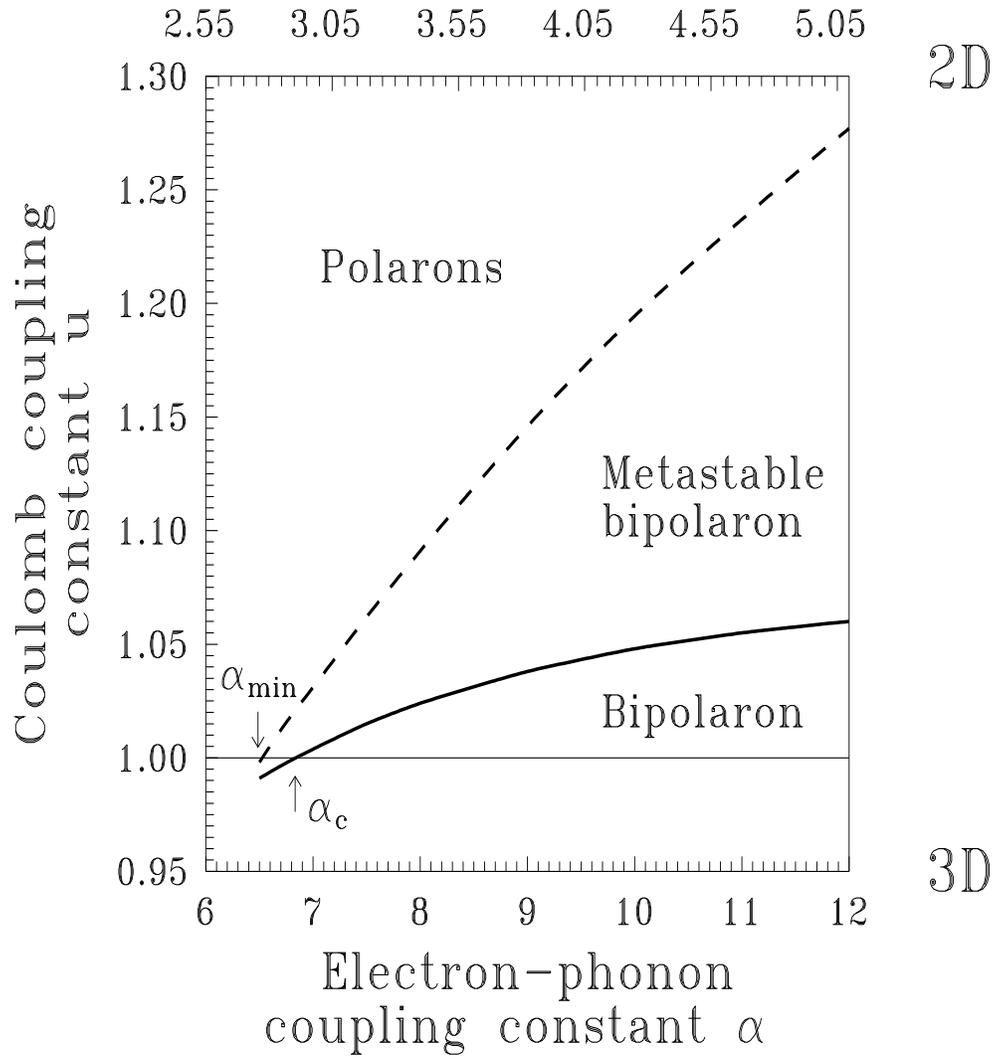,width=13cm,height=14cm}
\vspace*{1cm}
\caption{``Phase'' diagram in the plane of the coupling constants
($u,\alpha$).}
\label{fig2}\end{figure}

In Fig.~\ref{fig2} regions
are shown where polarons, bipolarons and metastable bipolaron states
exist. The top $x$-axis corresponds to the 2D case.
The metastable bipolaron is seen to occur for
$\alpha \geq \alpha_{min} \approx 6.5$ (in 3D)
which is slightly smaller than the critical value $\alpha_c \approx 6.8$.
Note also that a metastable bipolaron
was also reported in Ref. \onlinecite{iadon2}.

\begin{figure}
\hspace*{1cm}
\psfig{figure=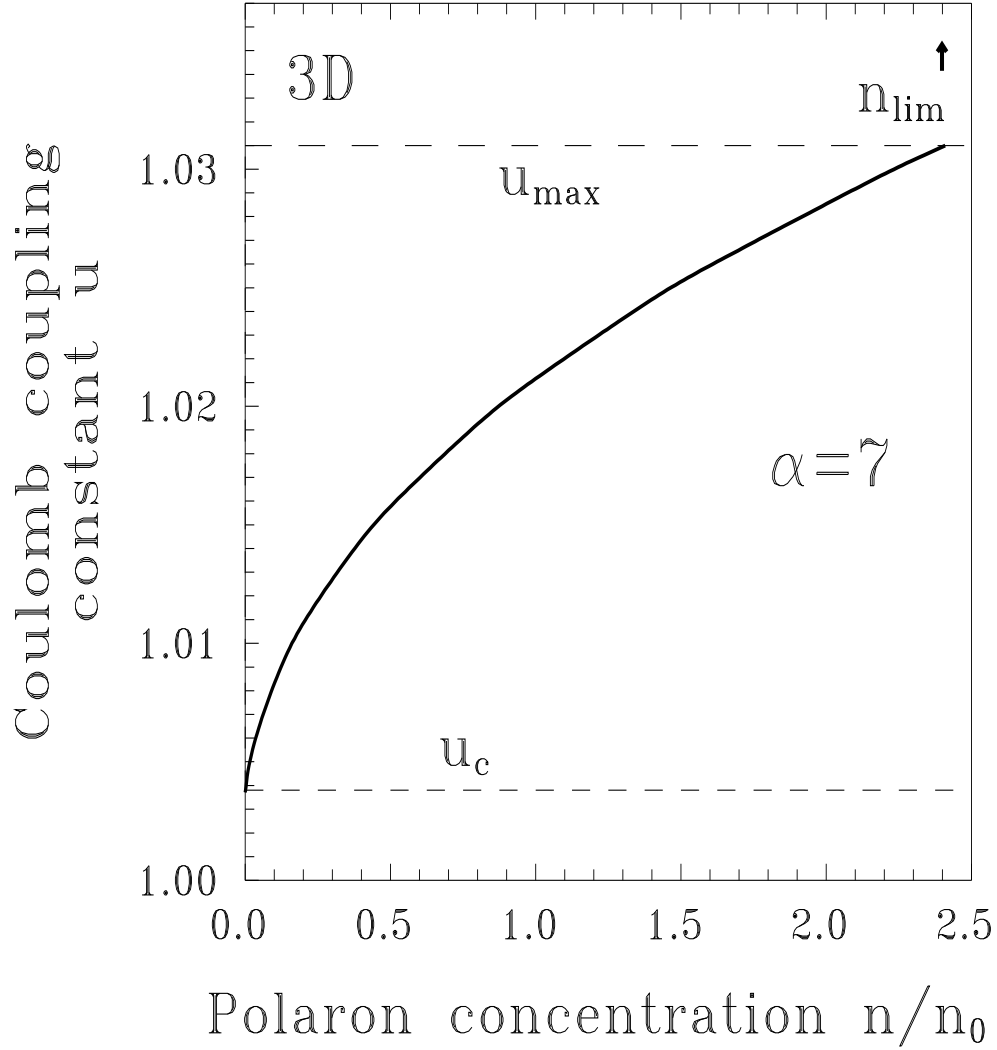,width=13cm,height=14cm}
\vspace*{1cm}
\caption{The limiting value of the Coulomb coupling constant $u$
vs. the polaron concentration for $\alpha =7$. The 3D metastable bipolaron
is stabilized by the polaron environment in the region below the shown curve.
The destabilization effect is negligible in this case.}
\label{fig3}\end{figure}

It follows from our stability criterion that
the metastable bipolaron is stabilized
by the polaron environment for concentrations $n> n_d$.
The stabilizing effect is evident for $\alpha$ close to the critical
value (we choose $\alpha=7$ for the 3D case
in Fig.~\ref{fig3}): the limiting value of
the Coulomb coupling constant increases with the polaron
concentration $n$. Bipolarons can exist
for the parameter values below the plotted curve. The
stability region of the isolated bipolaron is determined
by $u \leq u_c$. At $n=n_{lim} \approx 2.4n_0$ the constant $u$
reaches its maximal value $u_{max}$. Note that typical
concentrations for stability of bipolarons
$n_d \sim 10^{18}\div10^{19}\;{\rm cm}^{-3}$
are close to charge-carrier concentrations of heavily doped
degenerate polar semiconductors~\cite{wolf,varga,lemm}.

\begin{figure}
\hspace*{1cm}
\psfig{figure=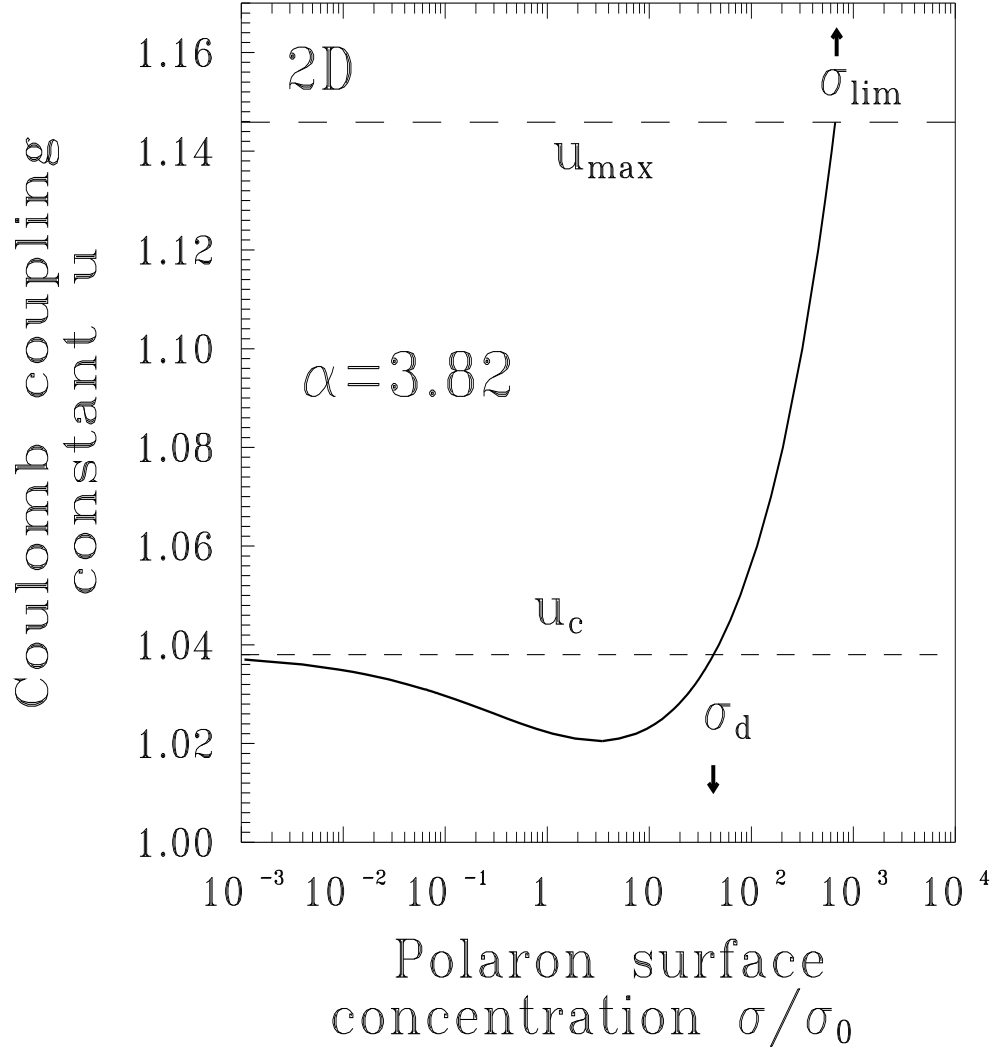,width=13cm,height=14cm}
\vspace*{1cm}
\caption{The limiting value of the Coulomb coupling constant $u$
vs. the polaron surface concentration for $\alpha =3.82$. The 2D
metastable bipolaron is stabilized by the polaron environment in the
region below the shown curve at $\sigma > \sigma_d \approx 43\sigma_0$.
The destabilization of the bipolaron
happens for the concentrations $0 < \sigma < \sigma_d$.}
\label{fig4}\end{figure}

The similar plot for $\alpha =3.82$ in 2D is presented in Fig. \ref{fig4}.
Both the stabilization (at $\sigma > \sigma_d \approx 43\sigma_0$)
and the destabilization (at $\sigma < \sigma_d$) effects are
seen clearly.

To conclude, we have formulated an in-medium stability criterion
for bipolarons which takes into account Fermi statistics
and the exchange energy. The former stablilize the bipolaron while
the latter destabilizes it. The competition of these two effects
depends on the polaron concentration. Out treatment of the bipolaron
is variational and therefore it is not to be excluded that
some of the relative minima found here for the bipolaron energy
as a function of variational parameters are artefacts of the method.

\acknowledgments
M.A.S. and A.A.S. thank the University of Antwerpen (UIA) for the
support and the kind hospitality during their visits to Belgium.
This work was supported by the F.W.O.-V project No. G.0287.95 and by
UA Bijzonder Onderzoeksfonds 1997 NOI "Thermodynamica van interagerende
identieke deeltjes met behulp van padintegralen'.

\end{document}